\documentclass[letterpaper]{article} % DO NOT CHANGE THIS
\usepackage{aaai2026}  % DO NOT CHANGE THIS
\usepackage{times}  % DO NOT CHANGE THIS
\usepackage{helvet}  % DO NOT CHANGE THIS
\usepackage{courier}  % DO NOT CHANGE THIS
\usepackage[hyphens]{url}  % DO NOT CHANGE THIS
\usepackage{graphicx} % DO NOT CHANGE THIS
\usepackage{natbib}  % DO NOT CHANGE THIS AND DO NOT ADD ANY OPTIONS TO IT
\usepackage{caption} % DO NOT CHANGE THIS AND DO NOT ADD ANY OPTIONS TO IT
\usepackage{algorithm}
\usepackage{algorithmic}
\usepackage{amsmath}
\usepackage{multicol}
\usepackage{multirow}
\usepackage{booktabs}
\usepackage{amssymb}
\usepackage{xcolor}
\usepackage{newfloat}
\usepackage{listings}
\DeclareCaptionStyle{ruled}{labelfont=normalfont,labelsep=colon,strut=off} % DO NOT CHANGE THIS
\floatstyle{ruled}
\newfloat{listing}{tb}{lst}{}
\floatname{listing}{Listing}
\title{GOMPSNR: Reflourish the Signal-to-Noise Ratio Metric \\ for Audio Generation Tasks}
\author{
    Lingling Dai\textsuperscript{\rm 1,\rm 2},
    Andong Li\textsuperscript{\rm 1,\rm 2}\thanks{Corresponding authors are Andong Li and Chengshi Zheng.},
    Cheng Chi\textsuperscript{\rm 1,\rm 2},
    Yifan Liang\textsuperscript{\rm 1,\rm 2},
    Xiaodong Li\textsuperscript{\rm 1,\rm 2},
    Chengshi Zheng\textsuperscript{\rm 1,\rm 2}\footnotemark[1]
}

\affiliations{
    \textsuperscript{\rm 1}Institute of Acoustics, Chinese Academy of Sciences, Beijing, China\\
    \textsuperscript{\rm 2}University of Chinese Academy of Sciences, Beijing, China\\
    \{dailingling, liandong, chicheng2021, liangyifan, lxd, cszheng\}@mail.ioa.ac.cn
}

\usepackage{bibentry}
\begin{document}

\maketitle

\begin{abstract}
In the field of audio generation, 
signal-to-noise ratio (SNR) has long served as an objective metric for evaluating audio quality. Nevertheless, recent studies have shown that SNR and its variants are not always highly correlated with human perception, prompting us to raise the questions: \textit{Why does SNR fail in measuring audio quality?} And \textit{how to improve its reliability as an objective metric?} In this paper, we identify the inadequate measurement of phase distance as a pivotal factor and propose to reformulate SNR with specially designed phase-distance terms, yielding an improved metric named GOMPSNR. We further extend the newly proposed formulation to derive two novel categories of loss function, corresponding to magnitude-guided phase refinement and joint magnitude-phase optimization, respectively. Besides, extensive experiments are conducted for an optimal combination of different loss functions. Experimental results on advanced neural vocoders demonstrate that our proposed GOMPSNR exhibits more reliable error measurement than SNR. Meanwhile, our proposed loss functions yield substantial improvements in model performance, and our well-chosen combination of different loss functions further optimizes the overall model capability. 
\end{abstract}

% Uncomment the following to link to your code, datasets, an extended version or similar.
% You must keep this block between (not within) the abstract and the main body of the paper.
\begin{links}
    \link{Code}{https://github.com/lingling-dai/GOMPSNR}
\end{links}

\section{Introduction}
In the development of audio technology, objective metrics have played a pivotal role in evaluating audio quality by providing standardized and quantitative measurements. Over the years, numerous metrics have been proposed, either with a reference (intrusive) or without a reference (non-intrusive). Among these metrics, the signal-to-noise ratio (SNR) or signal-to-distortion ratio (SDR) has been widely applied in audio signal processing and audio generation tasks, including speech enhancement (SE) \cite{fullsubnet}, bandwidth extension (BWE) \cite{eben,WSRGlow}, and blind speech separation (BSS) \cite{convtasnet}.

\begin{figure}[t]
\centering
\includegraphics[width=\columnwidth]{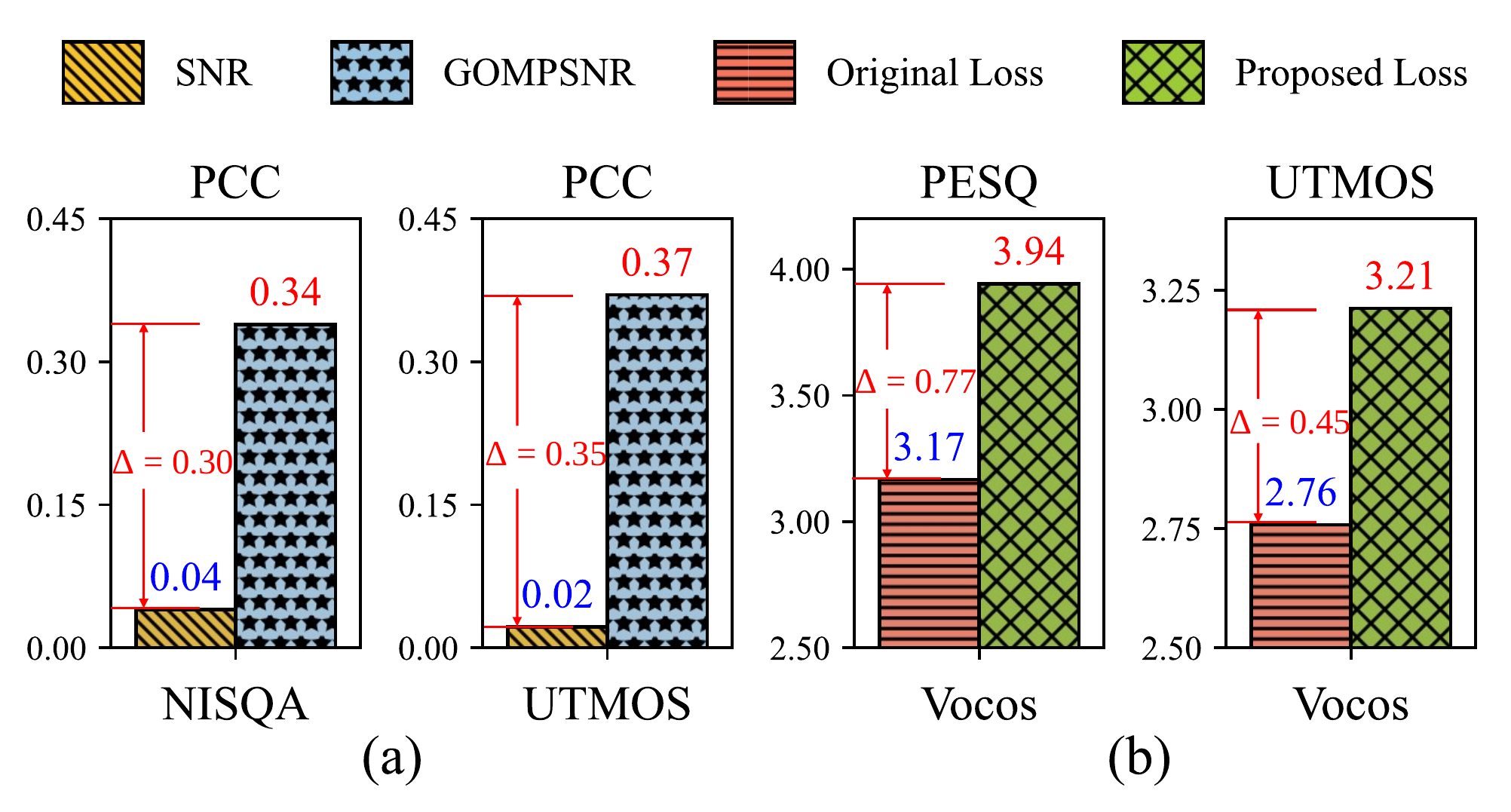} 
\caption{Illustration of the performance improvement brought by our proposed methods. (a) The correlation of SNR and our proposed GOMPSNR with other perceptual metrics. (b) The performance comparison between the original loss and our proposed loss. }
\label{fig:performance_illstration}
\end{figure}

Based on the strict mathematical formulation, SNR provides a straightforward measure of the difference between the estimated signal and the reference signal by computing the ratio of the power of the desired signal to that of the background noise or distortion. During the successive development of SNR, variants such as segmental SNR (segSNR) \cite{segsnr}, frequency weighted segSNR (\text{segSNR\textsubscript{fw}}) \cite{segSNR_fw}, and scale-invariant SNR (SI-SNR) \cite{SISDR} have been proposed to achieve more accurate and reliable measurements by taking into account the inherent characteristics of the audio signal. However, an increasing number of studies reveal that SNR and its variants may fail to provide consistent results with other perceptual metrics or subjective evaluations \cite{online-seanet,NU-GAN,lessons_urgent2024,TokenSplit,SELM}, and they are gradually being marginalized as a standard metric. Meanwhile, other metrics with similar mathematical formulations, such as Mel-cepstral distortion (MCD) \cite{mcd} and Multi-resolution short-time Fourier transform (M-STFT) \cite{mstft}, continue to remain mainstream. These findings lead us to raise the first question: \textbf{(i) Why does SNR fail in measuring the audio quality?}

To answer this question, we carry out detailed mathematical formula derivation and visualize the intermediate results to analyze the inherent characteristics of SNR. The results point to one possible factor, which is the inaccurate measurement of the distance of phase. Specifically, SNR can be seamlessly transformed from the time domain into the time-frequency (T-F) domain, where quantitative assessment of the raw waveform is transformed into the assessment of magnitude and phase in a coupled manner. As the phase spectrum exhibits a more irregular structure compared to the magnitude spectrum, measuring the distance of the phase component remains less dependable than that of the magnitude \cite{PHASEN}. The absence of an effective method for measuring the distance of phase is thus confirmed as the critical factor. Naturally, this leads to the second question: \textbf{(ii) How to improve the reliability of SNR as an objective metric?}

As the partial derivatives of the phase, \textit{i.e.}, instantaneous frequency (IF) and group delay (GD), exhibit clearer structures than the instantaneous phase (IP), calculating the distance of IF and GD between the estimated audio and the reference audio rather than IP can provide more reliable measurements \cite{online_dnn_differences}. Recent studies have leveraged the intrinsic properties of the phase derivatives for both phase optimization and phase discrepancies quantification. In \cite{rndvoc}, the phase derivatives are further developed from the omnidirectional perspective and consolidated into a more convenient and unified form. Based on that, we reformulate the SNR metric by replacing IP with the omnidirectional phase derivatives and correct the unstable terms in the formula, introducing \textbf{G}eneralized \textbf{OM}nidirectional \textbf{P}hase-oriented \textbf{S}NR, abbreviated as \textbf{GOMPSNR}, as an effective substitute metric of SNR. Meanwhile, we reinvestigate the formulation of several commonly adopted loss functions and propose novel forms of magnitude-guided phase refinement and joint magnitude–phase optimization by leveraging the property of phase derivatives. Experimental results on several state-of-the-art vocoders show that our proposed GOMPSNR exhibits higher relevance with perceptual metrics compared with SNR, and our reformulated loss functions significantly outperform their original loss functions. Additionally, comprehensive experiments are conducted to evaluate the optimal combination of different types of loss functions, and our well-chosen combination further improves the model performance.

\iffalse
Figure \ref{fig:performance_illstration} showcases the performance improvement brought by our proposed methods. 
 Our major contributions are summarized as follows:
 \begin{itemize}
     \item 
     \item 
     \item 
     \item 
 \end{itemize}

\fi

\section{Related Works}

\subsection{Audio Generation Tasks}
In recent years, the development of deep learning techniques has led to significant advancements in audio generation fields, which encompass a wide range of applications, such as text-to-speech (TTS) \cite{f5tts,cosyvoice}, voice conversion (VC) \cite{promptvc}, and singing voice synthesis (SVS) \cite{stylesinger}. Given an audio clip, text, or other modalities as input, these tasks aim to produce high-fidelity and diverse audio signals with high relevance, where generative models, including flow-based models, Generative Adversarial Networks (GANs), and diffusion probabilistic models are commonly employed \cite{flowtts,gradtts,VALL-E}. Despite the form of generative model varies, when designing the audio synthesis pipeline, a prevalent approach is to decompose the sophisticated audio generation process into several stages, with each focusing on a specific aspect of the task. Typically, a pretrained neural vocoder is employed to convert the mel-spectrogram or latent representations into a raw waveform in the last stage of such a paradigm, where the vocoder plays a crucial role in reconstructing the audio signals \cite{Vocos}.

\subsection{Objective Metrics for Audio Quality Assessment}
Despite subjective evaluation provides gold standards for evaluating the quality of audio signals, it is often laborious and time-consuming to collect massive subjective mean opinion scores (MOS). Moreover, MOS is sensitive to factors such as listener preferences and equipments. Instead, objective metrics provide quantitative assessments of audio quality and further facilitate fair comparisons between different methods from diverse acoustic aspects. One category of objective metrics measures raw waveform discrepancy or spectral distance in a point-wise manner, including SNR, Log Spectral Distance (LSD), and M-STFT. Moreover, some assess the audio quality by measuring the difference in terms of acoustic features. For instance, Periodicity Root Mean Square Error (RMSE), V/UV F1 score, and pitch RMSE \cite{f1_score} are developed to evaluate the periodicity, voicing, and pitch accuracy. Another category focuses on the perceptual quality, such as PESQ \cite{WB-PESQ}, UTMOS \cite{UTMOS}, and Scoreq \cite{scoreq_metric}, where the metrics are designed to align with human auditory perception and are often regarded as the economical and practical substitutes for subjective evaluations.

\subsection{Loss Functions for Enhancing the Audio Quality}
Loss functions are essential for guiding the training process and optimizing the model's performance, where the choice of loss function can significantly impact the quality of the generated audio. In the early stages of neural audio generation and signal processing development, the reconstruction losses were primarily composed of simple L1 or L2 losses of the raw waveform or the complex spectrum. However, these loss functions illustrate insufficient ability in enhancing the auditory quality, leading to further exploration of advanced loss functions. In \cite{hifigan}, a mel-spectrum loss is introduced for improving the perceptual quality inherent to the characteristics of the human auditory system. Similarly, \cite{PCS} designs a critical band importance function for perceptual enhancement of the training target. Furthermore, objective metrics and pretrained large-scale models are also integrated for model optimization \cite{train_obj,Finally}. Aside from the focus on perceptual loss design, several other studies concentrate on the point-wise optimization. For instance, \cite{NSPP} and \cite{rndvoc} propose phase-related loss functions to improve the continuity of phase spectra, which has been shown to further boost the audio quality.

\section{Methodology}
For audio signal processing and audio generation tasks, SNR and its variants have been previously widely applied in measuring the discrepancy between the estimated signal $\widehat{y}\in \mathbb{R}^n$ and the target $y\in \mathbb{R}^n$, where $n$ denotes the length of the raw waveform. By calculating the ratio of the power of the target signal to that of the residual error over the entire waveform, SNR reflects the energy discrepancy with a quite simple and intuitive mathematical formulation. However, recent researches have repeatedly confirmed that SNR illustrates a low correlation with auditory perception. To isolate the key factor of such misalignment, we shift the formulation from the time domain to the T-F domain, where the raw waveform is decomposed into the representation of magnitude and phase. We use $\{Y, \widehat{Y}\} \in \mathbb{C}^{L \times K}$, $\{ \theta, \widehat{\theta} \} \in \mathbb{R}^{L \times K}$ to present the complex spectrum and the phase spectrum of $\{y, \widehat{y}\}$, where the $L$ and $K$ denote the frame and frequency size, respectively. Figure \ref{fig:performance_illstration} illustrates the performance improvement brought by our proposed methods.

\begin{figure}[t]
\centering
\includegraphics[width=0.95\columnwidth]{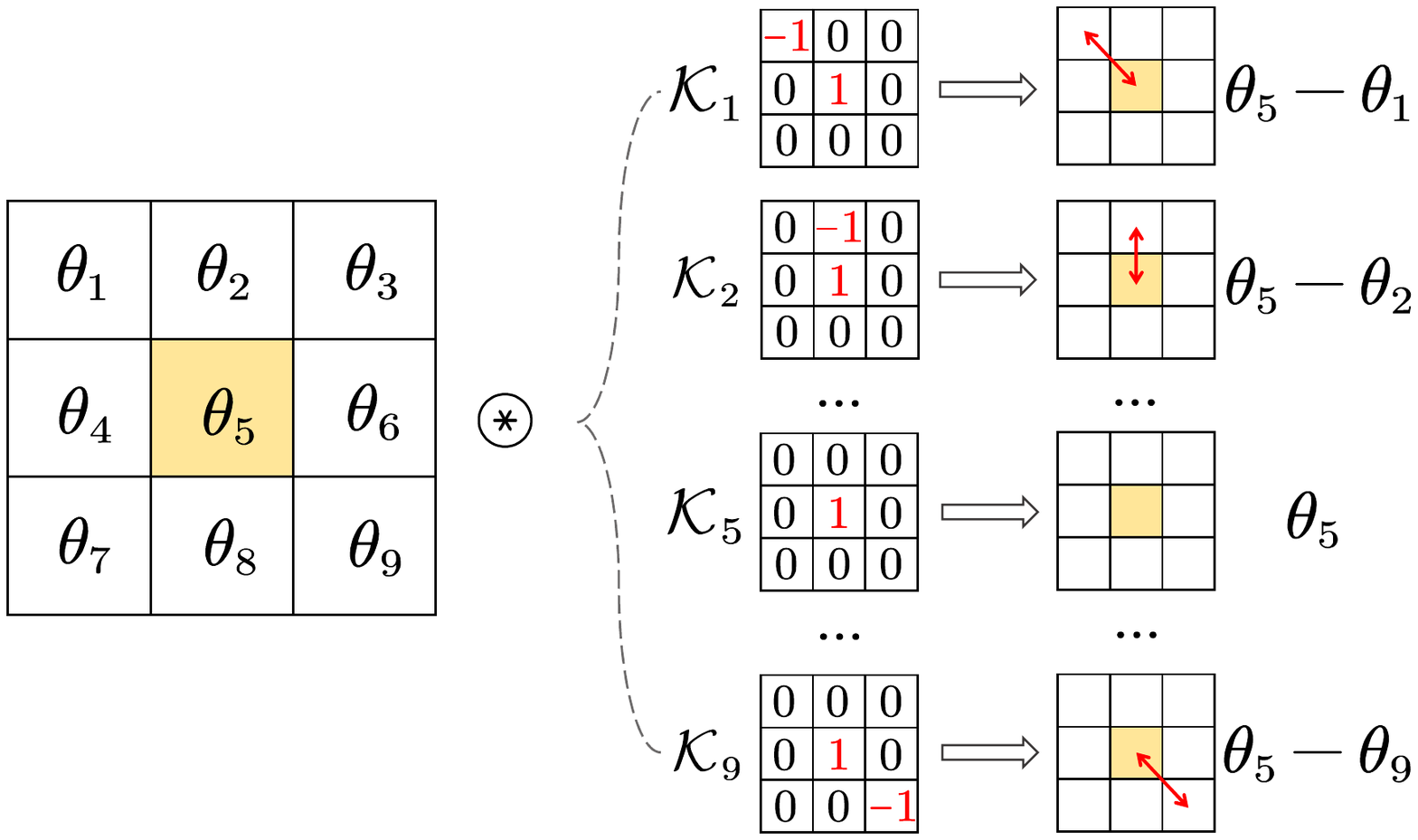} 
\caption{Illustration of obtaining the omnidirectional phase derivatives.}
\label{fig: phase_derivative}
\end{figure}

\subsection{Omnidirectional Phase Derivatives}
Compared with magnitude, the direct prediction of phase has remained a formidable challenge \cite{PHASEN}. Specifically, the phase spectrogram exhibits an irregular structure due to the wrapping property, which restricts the phase value to a limited range of $[-\pi, \pi)$. Moreover, phase is also highly sensitive to waveform shifts, which further brings challenges into the prediction process \cite{CMGAN_LOSS}. As computing phase derivatives and applying basic anti-wrapping functions yields more structural representations, researchers now design phase-aware loss functions and dedicated metrics to quantify phase discrepancies based on these unwrapped phase derivatives \cite{NSPP,SP-NSPP, choi2018phase_deepUnet,AP-BWE}. 

In \cite{rndvoc}, a novel omnidirectional phase (OP) representation is proposed to unify and complement the phase derivatives from omni directions. Specifically, as presented in Figure \ref{fig: phase_derivative}, nine $3\times 3$ kernels with fixed parameters $\mathcal{K}=Cat(\mathcal{K}_1, ..., \mathcal{K}_9)\in \mathbb{R}^{9\times 3 \times 3}$ are designed to obtain the derivatives from eight adjacent T-F bins, as well as the IP. Then the omnidirectional phase derivatives $\{ \nabla \theta, \nabla \widehat{\theta} \}$ are formulated as:
\begin{equation}
\nabla \theta=\theta \circledast \mathcal{K},\ \nabla \widehat{\theta}=\widehat{\theta}\circledast \mathcal{K},
\end{equation}
where $\circledast$ denotes the convolution operation. 

Moreover, an OP loss function is further derived to optimize the phase by minimizing the difference between the estimated phase and the target one:
\begin{equation}
\mathcal{L}_{OP}=\frac{1}{9KL} \sum_{i, k, l}  {f_{AW}\left( \nabla_i \theta - \nabla_i \widehat{\theta} \right)},
\label{eq:omni-phase}
\end{equation}
where an anti-wrapping function $f_{AW}$ is applied to unwrap the phase derivatives:
\begin{equation}
f_{AW}\left( x \right) =\left| x - 2\pi\cdot round\left( \frac{x}{2\pi} \right) \right|.
\label{eq:anti-wrapping}
\end{equation}

\begin{figure}[t]
\centering
\includegraphics[width=0.90\columnwidth]{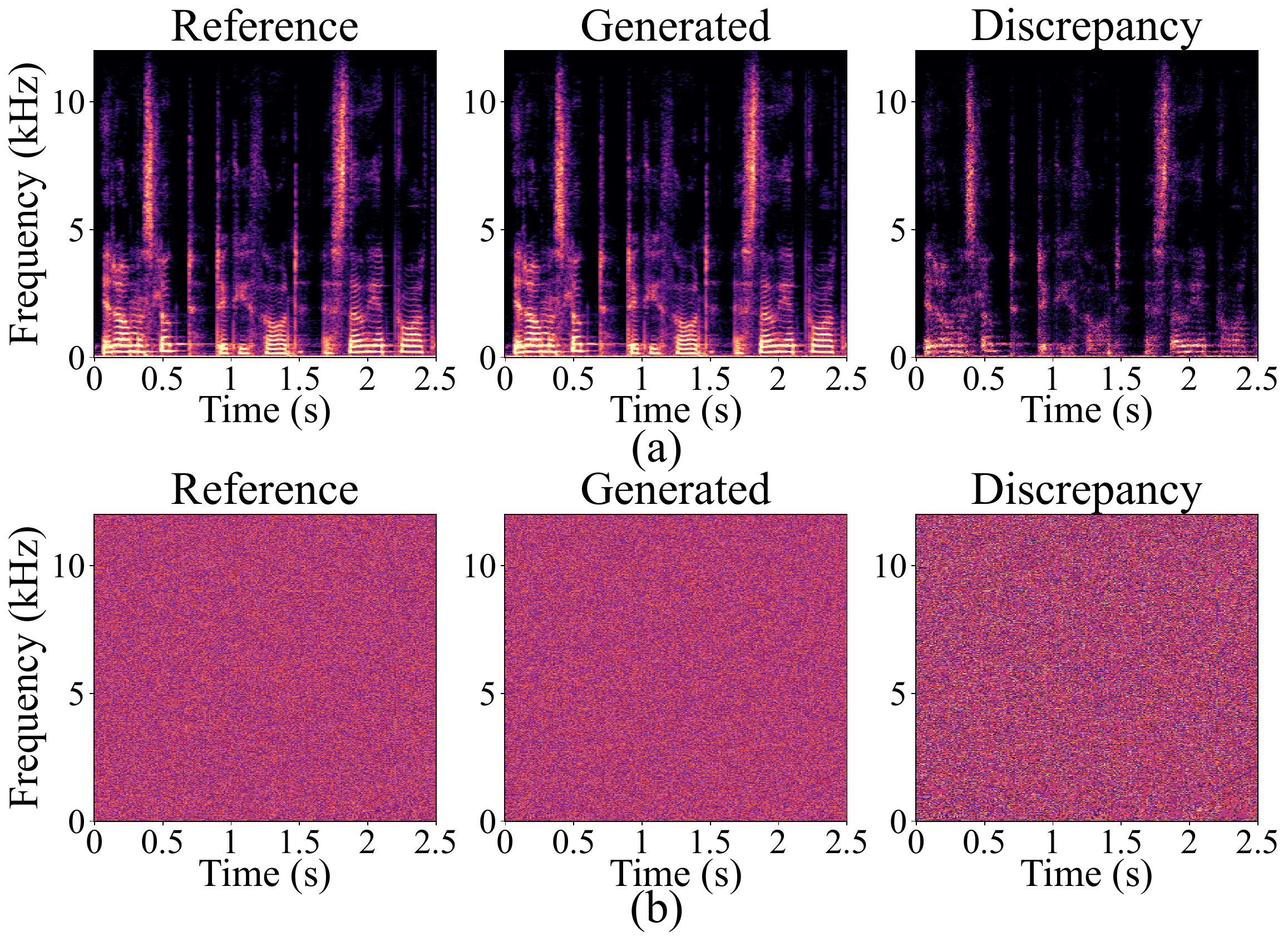} 
\caption{The visualization of the spectrogram discrepancy in terms of (a) magnitude and (b) phase.}
\label{fig: mag_pha_difference}
\end{figure}

 \begin{table}[]
\centering
\begin{tabular}{c|c}
\toprule
Metric & Definition \\ 
\midrule
SNR&   $-2|Y||\widehat{Y}|\cos \left( \theta -\widehat{\theta} \right)$ \\ 
OMPSNR&  $-\frac{2}{9}|Y||\widehat{Y}| \sum_{i} \cos \left( \nabla_i \theta -\nabla_i \widehat{\theta} \right)$   \\ 
GOMPSNR&  $\frac{2}{9}|Y||\widehat{Y}| \sum_{i} \left( \frac{1}{\pi}f_{AW}\left( \nabla_i\theta - \nabla_i\widehat{\theta} \right) -1 \right) $\\ 
\bottomrule
\end{tabular}%
\caption{The definition of the correlation component $C$ in SNR and our proposed OMPSNR and GOMPSNR. }
\label{tab:snr}
\end{table}

\begin{table*}[]
\centering
\small
\begin{tabular}{cc|cccccccc}
\toprule
  &  & &  &  & & V/UV& Periodicity & Pitch &   \\
Phase & RI & PESQ$\uparrow$    & UTMOS$\uparrow$ & MCD$\downarrow$  & M-STFT$\downarrow$  & F1$\uparrow$ & RMSE$\downarrow$  & RMSE$\downarrow$  & GOMPSNR$\uparrow$ \\ \midrule
-& -& 3.749 & 4.128 & 2.451 & 0.990 & 0.963 & 0.104 & 26.104 & 4.299 \\
P& -&3.711 & 4.082 & 2.500 & 0.994 & 0.962 & 0.107 & 25.323 & 4.282 \\
OP& -& 3.752 & 4.111 & 2.449 & 0.981 & 0.964 & 0.102 & 24.647 & 4.395 \\
WOP& -& \textbf{3.928}& \textbf{4.168}& \textbf{2.256}& \textbf{0.964}& \textbf{0.969}& \textbf{0.088}& \textbf{20.698}& \textbf{5.232} \\ \midrule

WOP& RI (L1)& 3.891 & 4.180 & 2.394 & 1.015 & 0.966 & 0.098       & 19.557     & 5.718             \\
WOP& ORI (L2)  & 3.791 & 4.123 & 2.614 & 1.051 & 0.965 & 0.097       & 19.488     & 5.593             \\
WOP& ORI (L1)  & 3.998 & \textbf{4.207}& 2.218 & 0.957 & \textbf{0.971}& \textbf{0.083}& \textbf{19.138}& \textbf{5.818}\\
WOP& CORI (L2) & \textbf{4.001}& 4.164 & 2.238 & \textbf{0.935}& \textbf{0.971}& \textbf{0.083}& 19.176     & 5.674             \\
WOP& CORI (L1) & 3.992 & 4.186 & \textbf{2.212}& 0.944 & \textbf{0.971}& 0.085       & 19.476     & 5.622 \\           
\bottomrule
\end{tabular}%
\caption{Experimental results of phase-oriented loss functions and co-optimized phase and magnitude loss functions on the LJSpeech Dataset. Inside the parentheses is the adopted type of the point-wise distance. The \textbf{optimal} results are marked in \textbf{bold}. }
\label{tab:ri}
\end{table*}

\subsection{Generalized Omnidirectional Phase-oriented SNR}
In the T-F domain, SNR is formulated as the ratio of two components, where the numerator measures the energy of the target spectrogram, and the denominator aggregates the total squared deviation between the target and the estimated spectrograms:
\begin{equation}
SNR =10\log_{10} \frac{\sum_{k,l}{|Y|^2}}{\sum_{k,l}{|Y-\widehat{Y}|^2}}. \\
\end{equation}
The denominator can be unfolded into three additive components: the energy of the target spectrogram, the energy of the estimated spectrogram, and a summation of the signed correlation component $C$:   
 \begin{equation}
SNR = 10\log_{10} \frac{\sum_{k,l}{|Y|^2}}{\sum_{k,l}{\left( |Y|^2+|\widehat{Y}|^2 + C \right) }},
 \label{eq:snr}
\end{equation}
where $C$ is presented in Table \ref{tab:snr}.

\begin{figure}[t]
\centering
\includegraphics[width=1.0\columnwidth]{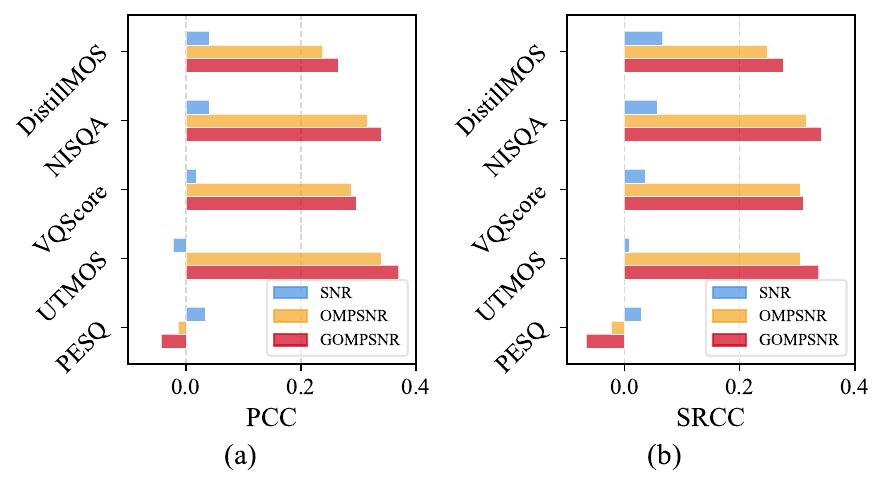} 
\caption{The correlation of SNR and GOMPSNR with several commly adopted perceptual metrics in terms of PCC and SRCC on the LibriTTS dataset.}
\label{fig:snr}
\end{figure}

In the above equation, SNR is further transformed into a coupled manner of magnitude and phase. To disentangle the influence of each component, we visualize the residuals of the magnitude and phase spectrogram between the vocoder output and its ground-truth reference in Figure \ref{fig: mag_pha_difference}. As one can observe, the discrepancy of magnitude spectrogram is structurally salient and clearly highlights the regions where the model fails to recover the spectral details. Instead, the visualization of the phase discrepancy yields uninformative results, indicating that conventional metrics, whether explicitly or implicitly calculating the phase distance, may have miscalculated the phase discrepancy. To mitigate the deviation in measuring the distance of phase, we propose to use the omnidirectional phase derivatives to replace the IP and upgrade SNR into \textbf{OM}nidirectional \textbf{P}hase-oriented \textbf{SNR}, dubbed \textbf{OMPSNR}, that simultaneously accounts for the magnitude discrepancy and the misalignment of omnidirectional phase derivatives. 

Moreover, we find that the sign of the correlation component $C$ changes around the value of $\pm \pi /2$ for $\theta - \widehat{\theta}$, which probably leads to the numerical oscillation in the summation of $C$. Such a pattern intensifies the influence of phase, making SNR highly sensitive to perturbations in phase errors. Therefore, we further propose the generalized version of OMPSNR (\textbf{GOMPSNR}), by modifying the correlation component $C$ into a non-positive item to alleviate the instability caused by the phase error. Additionally, the nonlinear transformation is replaced with a strictly linear mapping function. The final formulation is presented in Table \ref{tab:snr}.

% Please add the following required packages to your document preamble:
% \usepackage{booktabs}
% \usepackage{multirow}
% \usepackage{graphicx}
\begin{table*}[]
\centering
\small
\begin{tabular}{c|ccc|cccccccc}
\toprule
                        \multirow{2}{*}{Model}&      &   &     & &  &  &  & V/UV & Periodicity & Pitch &  \\
                         & Mag    & Phase & RI   & PESQ$\uparrow $   & UTMOS$\uparrow $& MCD$\downarrow $ & M-STFT$\downarrow $ & F1$\uparrow $& RMSE$\downarrow $ & RMSE$\downarrow $ & GOMPSNR$\uparrow $ \\ \midrule
\multirow{5}{*}{Vocos}  & -      & -     & -    & 3.749 & 4.128 & 2.451 & 0.990 & 0.963 & 0.104       & 26.104    & 4.299   \\
                        & Lin & -     & -    & 3.775 & 4.096 & 2.428 & 0.972 & 0.965 & 0.099       & 23.975    & 4.482   \\
                        & -      & WOP   & -    & 3.928 & 4.168 & 2.500 & 0.964 & 0.969 & 0.088       & 20.698    & 5.232   \\
                        & -      & -     & CORI & 3.806 & 4.146 & 2.424 & 0.963 & 0.964 & 0.101       & 24.238    & 4.653   \\
                        & Lin & WOP   & CORI & \textbf{4.035}& \textbf{4.190}& \textbf{2.162}& \textbf{0.922}& \textbf{0.971}& \textbf{0.082}& \textbf{19.657}& \textbf{5.749}\\ \midrule
\multirow{5}{*}{APNet}  & Log    & P     & RI   & 2.836 & 2.950 & 3.716 & 1.134 & 0.952 & 0.132       & 27.502    & 3.335   \\
                        & Lin & P     & RI   & 3.124 & 3.235 & 5.924 & 1.437 & 0.936 & 0.153       & 26.278    & -0.957  \\
                        & Log    & WOP   & RI   & 3.503 & 3.648 & \textbf{3.190}& \textbf{1.032}& \textbf{0.959}& \textbf{0.112}& \textbf{22.310}& \textbf{4.318}\\
                        & Log    & P     & CORI & 3.293 & 3.211 & 3.696 & 1.083 & 0.953 & 0.129       & 26.656    & 2.674   \\
                        & Lin & WOP   & CORI & \textbf{3.569}& \textbf{3.676}& 7.144 & 1.461 & 0.936 & 0.149       & 23.811    & -3.381  \\ \midrule
\multirow{5}{*}{APNet2} & Log    & P     & RI   & 3.643 & 3.764 & 2.769 & 1.041 & 0.960 & 0.111       & 22.812    & 4.961   \\
                        & Lin & P     & RI   & 3.731 & 3.844 & 2.667 & 1.034 & 0.964 & 0.100       & 22.490    & 5.163   \\
                        & Log    & WOP   & RI   & 3.803 & \textbf{4.073}& 2.518 & 1.002 & 0.967 & 0.093       & \textbf{19.543}& \textbf{5.629}\\
                        & Log    & P     & CORI & 3.702 & 3.784 & 2.596 & 0.972 & 0.964 & 0.102       & 23.055    & 4.889   \\
                        & Lin & WOP   & CORI & \textbf{3.901}& 4.056 & \textbf{2.369}& \textbf{0.949}& \textbf{0.970}& \textbf{0.086}& 19.729    & 5.533   \\ \midrule
\multirow{5}{*}{RNDVoc} & Log    & P     & RI   & 4.033 & 4.085 & 1.946 & 0.877 & 0.974 & 0.080       & 19.241    & 5.655   \\
                        & Lin & P     & RI   & 4.029 & 4.047 & 1.959 & 0.910 & 0.973 & 0.082       & 19.236    & 5.636   \\
                        & Log    & WOP   & RI   & 4.105 & \textbf{4.231}& 1.872 & 0.884 & 0.974 & 0.076       & \textbf{18.000}& \textbf{5.870}\\
                        & Log   & P     & CORI & 4.064 & 4.135 & 1.896 & 0.871 & 0.973 & 0.079       & 19.173    & 5.588   \\
                        & Lin & WOP   & CORI & \textbf{4.121}& 4.223 & \textbf{1.844}& \textbf{0.875}& \textbf{0.975}& \textbf{0.074}& 18.026    & 5.822  \\ \bottomrule
\end{tabular}%
\caption{Objective results of different vocoder models with various combinations of loss function  on the LJSpeech Dataset.}
\label{tab:ljspeech}
\end{table*}

\subsection{Coupling Magnitude and Phase in Loss Functions}
Apart from metrics that suffer from the assessment deviation brought by the incorrect measurement of phase distance, existing loss functions may also face the inherent difficulties of implicit or direct phase optimization. While the OP loss has been proven effective \cite{rndvoc}, we further explore more generalized designs of loss function that cooperate phase derivatives and magnitude to innovate the classical ones. Specifically, we reformulate the loss function from two different aspects: magnitude-guided phase refinement and joint magnitude–phase optimization. 

Firstly, following \cite{CMGAN_LOSS,bapen}, we apply a magnitude-based weighting term to the OP loss function, assigning region-specific importance according to the relative value of the magnitude spectrogram. The weighted OP (WOP) loss is defined as:

\begin{equation}
\mathcal{L}_{WOP}=\frac{1}{9KL} \sum_{i, k, l} \frac{|Y|\cdot {f_{AW}\left( \nabla_i \theta - \nabla_i \widehat{\theta} \right)}}{\text{max}(|Y|)}.
\label{eq:w-omni-phase}
\end{equation}

Then we reformulate the commonly adopted RI loss, which shares a similar formulation with SNR. Aligned with OMPSNR, one straightforward operation is to substitute the IP with the omnidirectional phase derivatives. Then we propose the OmniRI (ORI) loss, given by: 

\begin{equation}
\begin{aligned}
\mathcal{L}_{ORI}&=\frac{1}{9}\sum_{i}h\left( |Y|\cos \big( \nabla _i \theta \big) ,|\widehat{Y}|\cos \big( \nabla _i \widehat{\theta} \big) \right) \\&+ \frac{1}{9}\sum_{i}h\left( |Y|\sin \big( \nabla _i \theta \big) ,|\widehat{Y}|\sin \big( \nabla _i \widehat{\theta} \big) \right) ,
\end{aligned}
\end{equation}
where $h(\cdot, \cdot)$ denotes the point-wise distance, \textit{i.e.}, L1 and L2 distance. Note that the point-wise averaging operator is omitted for clarity.  
Additionally, we further combine the phase derivatives and the magnitude in a coupled manner to obtain the Coupled OmniRI (CORI) loss:
\begin{equation}
\mathcal{L}_{CORI}=\frac{2}{9\pi}\sum_{i} h\left( |Y|,\ |\widehat{Y}| \right) f_{AW}\left( \nabla _i \theta -\nabla _i \widehat{\theta} \right).
\end{equation}

\begin{table*}[]
\centering
\small
\begin{tabular}{c|ccc|cccccccc}
\toprule
                        \multirow{2}{*}{Model}&      &   &     & &  &  &  & V/UV & Periodicity & Pitch &  \\
                         & Mag    & Phase & RI   & PESQ$\uparrow $   & UTMOS$\uparrow $& MCD$\downarrow $ & M-STFT$\downarrow $ & F1$\uparrow $& RMSE$\downarrow $ & RMSE$\downarrow $ & GOMPSNR$\uparrow $ \\ \midrule
\multirow{5}{*}{Vocos }  & -      & -   & -    & 3.167 & 2.758 & 3.928 & 1.074 & 0.920 & 0.166 & 43.456  & 3.909 \\
                        & Lin& -   & -    & 3.298 & 2.910 & 3.641 & 0.995 & 0.940 & 0.136 & 33.961  & 4.314 \\
                        & -      & WOP & -    & 3.886 & 3.120 & 3.028 & 0.927 & 0.957 & 0.103 & 27.835  & 5.306 \\
                        & -      & -   & CORI & 3.227 & 2.830 & 3.752 & 1.010 & 0.929 & 0.148 & 38.231  & 4.122 \\
                        & Lin& WOP & CORI & \textbf{3.942}& \textbf{3.212}& \textbf{2.866}& \textbf{0.887}& \textbf{0.959}& \textbf{0.101}& \textbf{22.585}& \textbf{5.777}\\  \midrule
\multirow{5}{*}{APNet}  & Log    & P   & RI   & 2.676 & 2.034 & 4.453 & 1.307 & 0.926 & 0.159 & 42.059  & 3.532 \\
                        & Lin& P   & RI   & 3.030 & 2.376 & 4.018 & 1.241 & 0.932 & 0.152 & 43.897  & 4.262 \\
                        & Log    & WOP & RI   & 3.288 & 2.536 & 3.837 & 1.144 & 0.946 & 0.128 & \textbf{22.879}& 4.792 \\
                        & Log    & P   & CORI & 3.083 & 2.292 & 3.819 & 1.074 & 0.938 & 0.142 & 38.061  & 4.048 \\
                        & Lin& WOP & CORI & \textbf{3.654}& \textbf{2.725}& \textbf{3.344}& \textbf{0.915}& \textbf{0.957}& \textbf{0.105}& 27.180  & \textbf{5.297}\\  \midrule
\multirow{5}{*}{APNet2 } & Log    & P   & RI   & 1.685 & 1.327 & 6.305 & 1.998 & 0.707 & 0.360 & 220.525 & 1.615 \\
                        & Lin& P   & RI   & 1.819 & 1.356 & 5.728 & 1.777 & 0.772 & 0.320 & 230.402 & 1.969 \\
                        & Log    & WOP & RI   & 2.932 & 2.454 & 4.110 & 1.162 & 0.934 & 0.146 & 33.382  & 4.343 \\
                        & Log    & P   & CORI & 3.347 & 2.605 & 3.607 & 0.978 & 0.946 & 0.125 & 27.798  & 4.532 \\
                        & Lin& WOP & CORI & \textbf{3.789}& \textbf{2.974}& \textbf{3.187}& \textbf{0.901}& \textbf{0.957}& \textbf{0.107}& \textbf{21.601}& \textbf{5.269}\\  \midrule
\multirow{5}{*}{RNDVoc } & Log    & P   & RI   & 4.071 & 3.106 & 2.347 & 0.779 & 0.964 & 0.088 & 36.525  & 6.056 \\
                        & Lin& P   & RI   & 4.103 & 3.175 & 2.253 & 0.775 & 0.966 & 0.085 & 26.900  & 6.203 \\
                        & Log    & WOP & RI   & \textbf{4.162}& 3.285 & \textbf{2.203}& \textbf{0.764}& 0.968 & 0.080 & 26.540  & \textbf{6.646}\\
                        & Log    & P   & CORI & 4.102 & 3.200 & 2.293 & 0.774 & 0.967 & 0.085 & \textbf{25.916}& 6.051 \\
                        & Lin& WOP & CORI & 4.159 & \textbf{3.291}& 2.237 & 0.767 & \textbf{0.969}& 0.079& 28.257  & 6.559 \\  \bottomrule
\end{tabular}%
\caption{Objective results of different vocoder models with various combinations of loss function on the LibriTTS Dataset.}
\label{tab:libritts}
\end{table*}

\section{Implementation Details}
\subsection{Data Preparation}
We conduct experiments on two commonly adopted benchmarks in neural vocoders, namely LJSpeech \cite{ljspeech} and LibriTTS \cite{libritts}. The LJSpeech dataset is a relatively small benchmark with 13,100 clean speech clips by a single female speaker, and the sampling rate is 22.05 kHz. Aligned with previous works, we use the division in the open-sourced VITS repository\footnote{https://github.com/jaywalnut310/vits/tree/main/filelists} for training, validation, and testing, respectively. The LibriTTS dataset contains approximately 960 hours of speech with a sampling rate of 24 kHz. Following \cite{bigvgan}, we use the entire training subsets (train-clean and train-other) for training and adopt the subsets test-clean and test-other for testing. For LJSpeech, the mel-spectrogram dimension is set to 80, with a hop size of 256 and a window size of 1024, and an effective frequency range from 0 to 8 kHz. For LibriTTS, the mel-spectrogram dimension is set to 100 at a full frequency range of 0 to 12 kHz, with identical hop size and window size settings as LJSpeech. Additionally, we use a hanning window with a window size of 1024 and a hop size of 256 to transform the raw waveform into the complex spectrum when calculating GOMPSNR.

\subsection{Training Settings}
In this paper, several state-of-the-art neural vocoders, including Vocos \cite{Vocos}, APNet \cite{ai2023apnet}, APNet2 \cite{apnet2}, and RNDVoc \cite{rndvoc}, are implemented in the experiments. For training, we follow the official training pipeline of APNet2\footnote{https://github.com/redmist328/APNet2} for all vocoders, except that the learning rate is set to 5e-4. Specifically, the multi-period discriminator (MPD) \cite{hifigan} and multi-resolution spectrogram discriminator (MRSD) \cite{univnet} are adopted for adversarial training, with the hinge GAN losses used as the adversarial training loss. Additionally, the feature matching loss and the mel-spectrogram loss are also included as the basic loss functions. We retain all other loss functions adopted in the original paper for each vocoder and each is trained for 2 million steps. \footnote{Note that the results may be different from the original papers due to different training settings.} Additionally, Neural Audio Codecs (NACs) including WavTokenizer \cite{ji2024wavtokenizer} and Vocos \cite{Vocos} are implemented for auxiliary evaluations. We adopt the official implementations of Vocos\footnote{https://github.com/gemelo-ai/vocos} and WavTokenizer\footnote{https://github.com/jishengpeng/WavTokenizer}. Following \cite{Vocos}, we only optimize the decoder with frozen encoder and codebooks. Models are trained on the LibriTTS dataset for 2 million steps.

\section{Results and Discussion}
\subsection{Validation on GOMPSNR}
In this section, we validate the performance of OMPSNR and GOMPSNR as an intrusive metric in comparison with SNR. Several perceptual objective metrics, including PESQ \cite{WB-PESQ}, UTMOS \cite{UTMOS}, VQScore \cite{vqscore}, NISQA \cite{nisqa}, and DistillMOS \cite{nisqa}, are employed as perceptual references for calculating the Pearson Correlation Coefficient (PCC) and the Spearman’s Rank Correlation Coefficient (SRCC), with higher absolute values indicating stronger relevance. Figure \ref{fig:snr} presents the results of the officially pretrained Vocos on LibriTTS. As one can observe, SNR exhibits weak correlation with all perceptual metrics, with PCC and SRCC scores not exceeding 0.1, suggesting its inefficacy as an objective metric. In contrast, by simply substituting the IP counterpart with omnidirectional phase derivatives in SNR, the correlation with these perceptual metrics improves significantly, further confirming that the inaccurate estimation of phase distance is a primary reason for the diminished efficacy of SNR. Moreover, the rectification of the correlation component in GOMPSNR further improves the effectiveness, where GOMPSNR outperforms OMPSNR and exhibits a comparatively strong correlation with most of the perceptual metrics in terms of PCC and SRCC, indicating its superior performance as an intrusive metric for audio quality assessment. With a straightforward and interpretable mathematical formulation, GOMPSNR is convenient to be applied in any audio generation or signal processing tasks that require frame-wise alignment and is expected to become a new standard metric for audio quality assessment.

\subsection{Phase-oriented Loss Functions}
As the under-modeling of phase distance leads to insufficient modeling of SNR as an objective metric, the incorrect phase optimization in training process may also hinder the overall model performance. In this section, we evaluate the impact of phase-oriented loss functions on audio generation tasks by conducting experiments on Vocos. The objective results are presented in the upper bound of Table \ref{tab:ri}. Note that we use P to denote the vanilla phase loss proposed in \cite{NSPP}. As illustrated in Table \ref{tab:ri}, trained with hinge GAN loss, feature matching loss, and mel-spectrogram loss, Vocos is able to achieve satisfying performance on the LJSpeech dataset, and the two other phase-oriented loss functions including the vanilla phase loss and the OP loss provide no further improvement. However, the proposed WOP loss, which is a simple magnitude-weighted version of the OP loss, brings significant improvements to Vocos across all objective metrics. Such finding highlights the benefits of integrating magnitude as prior information and further underlines the need to investigate how phase and magnitude should be coupled during the phase optimization process.

\subsection{Joint Magnitude and Phase Optimization}
In addition to incorporating magnitude as an auxiliary guidance in phase-oriented loss functions, another common practice in loss function design is to explore the joint optimization of magnitude and phase. To validate whether the co-optimized magnitude and phase can lead to better performance, we conduct experiments on Vocos with different RI losses. The results are given in the lower bound of Table \ref{tab:ri}, and we use RI to denote the vanilla RI loss. It is noticeable that the vanilla RI loss even slightly degrades the model performance on most of the objective metrics. In contrast, our proposed OmniRI loss and the Coupled OmniRI loss in L1 distance further enhance the model performance, even assisted with the WOP loss that provides direct supervision on phase. These results indicate that disentangling the explicit phase component from the complex spectrum and reforming it benefits both magnitude and phase retrieval, despite the magnitude counterpart remaining unchanged. Furthermore, co-optimizing the phase and magnitude with properly designed forms is demonstrated to further enhance the overall performance and the Coupled OmniRI loss exhibits robust performance on point-wise distance choice.

\subsection{Comprehensive Evaluation on Loss Functions}
In this section, we conduct comprehensive experiments on advanced vocoders to validate an optimal combination of different loss functions, covering magnitude-oriented, phase-oriented, and the co-optimized form of both magnitude and phase, by replacing the corresponding losses in each vocoder with our well-chosen alternatives. Table \ref{tab:ljspeech} and Table \ref{tab:libritts} present the results of different vocoders on the LJSpeech dataset and LibriTTS dataset, respectively. Note that we use the Coupled OmniRI loss in L1 distance by default. Remarkably, each alternative loss function outperforms the original one used in the paper, demonstrating that the original loss function did not fully exploit the model's potential. One may notice that the linear form of the magnitude loss leads to the degradation on the LJSpeech dataset in terms of MCD, M-STFT, and GOMPSNR, which can likely be attributed to the overfitting of speech energy on such a limited dataset. However, it is more effective in improving the perceptual quality compared with the logarithmic one. Furthermore, the combination of our selected loss functions consistently outperforms the original loss function settings across all the vocoders, including RNDVoc, which already achieves outstanding results without elaborately designed losses. The experimental results further reveal that poorly designed loss functions tend to yield sub-optimal performance, whereas our proposed loss functions exploit the model’s full potential and significantly enhance the audio quality. Meanwhile, our proposed GOMPSNR also  exhibits a strong correlation with other metrics, indicating its effectiveness for providing a quantitative assessment between the estimated signal and the reference.

\subsection{Evaluation on Neural Audio Codecs}
\begin{figure}[t]
\centering
\includegraphics[width=0.98\columnwidth]{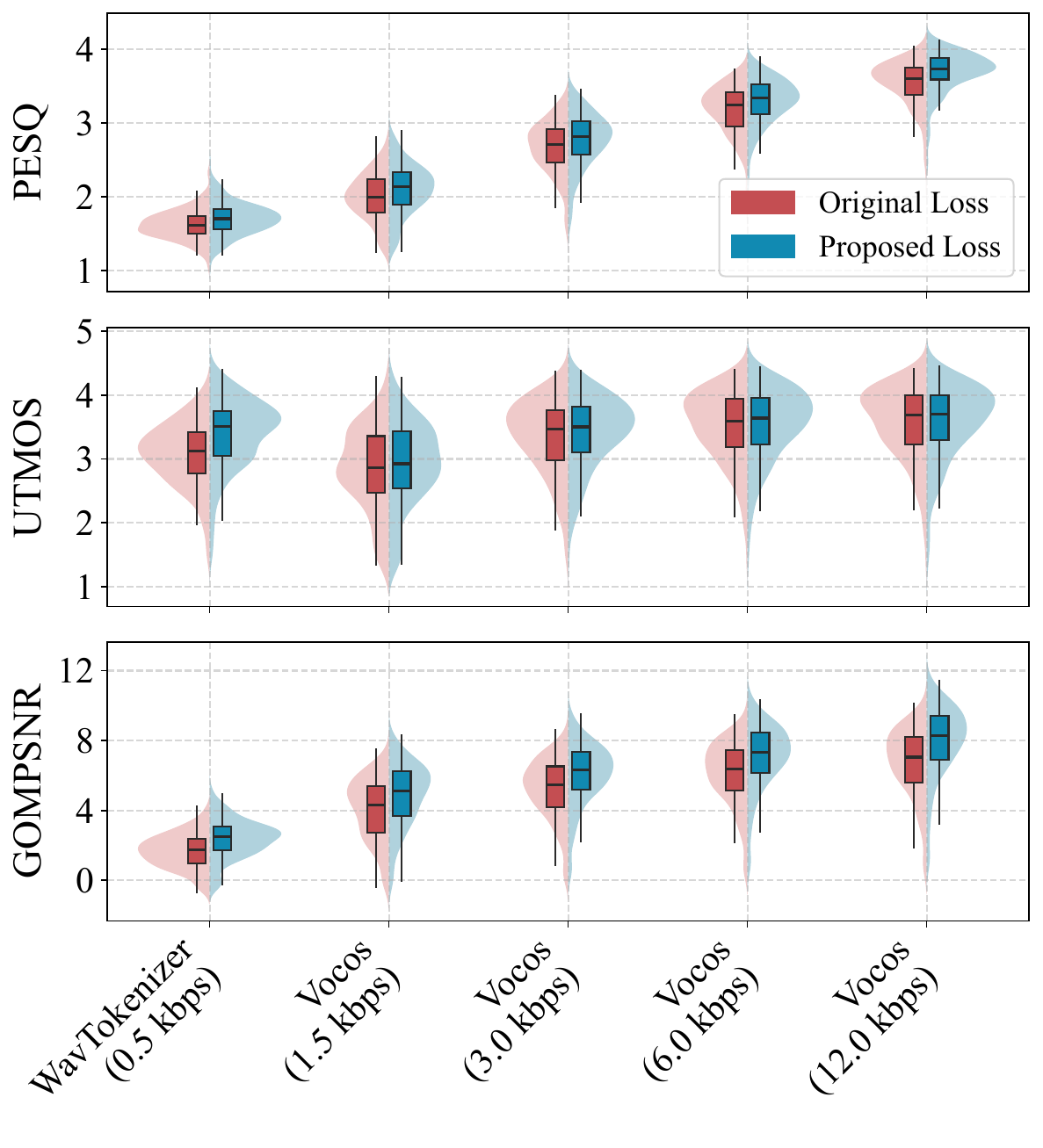} 
\caption{The objective results of different codec models on the LibriTTS Dataset.}
\label{fig:codec}
\end{figure}

To verify the applicability of our well-chosen combination of loss functions on other audio generation tasks, we further conduct extensive experiments on NACs, which compress the raw waveform into discrete acoustic codec representations and then reconstruct the audio from the latent representations.
As presented in Figure \ref{fig:codec}, after employing our well-chosen losses, both WavTokenizer and Vocos demonstrate superior performance across different bandwidths compared with their original configurations. Notably, the improvement is more pronounced at lower bandwidths, indicating that this set of loss functions is especially beneficial under higher compression levels. The appealing results further confirm the effectiveness of our proposed loss functions in enhancing the audio quality in audio generation tasks, and also demonstrate the applicability of our proposed GOMPSNR as an objective metric for evaluating the audio quality.

\section{Conclusion}
This paper proposes GOMPSNR, an effective alternative to SNR as an objective metric. By reformulating SNR with modified phase-distance terms, GOMPSNR exhibits improved measurement rationality and a stronger correlation with auditory perception. Furthermore, we extend GOMPSNR to develop novel loss functions, including a phase-oriented type and a co-optimized form of magnitude and phase, and explore a feasible combination of different loss functions. Quantitative experiments validate the effectiveness of GOMPSNR in audio generation tasks, highlighting its potential as a reliable objective metric for evaluating audio quality. Additionally, extensive experimental results demonstrate that our proposed loss functions yield significant performance improvement on advanced neural vocoders, and our well-chosen combination of loss functions manages to further enhance the overall audio quality. 

\section*{Acknowledgments}\label{sec:acknowledgement}
This work was supported by the National Natural Science Foundation of China (NSFC) under Grants 62501588.

\bibliography{aaai2026}

\end{document}